\documentstyle[amssymb,12pt,epsfig]{article}

\begin{document}

{\large Multiple giant resonances in nuclei: their excitation and
decay\bigskip }

M. S. Hussein$^{1}$, B. V. Carlson$^{2}$, and L. F. Canto$^{3}\bigskip $

$^{1}$Instituto de Fisica, Universidade de S\~{a}o Paulo,

C.P. 66318, 05315-970, S\~{a}o Paulo, SP, Brazil\medskip

$^{2}$Departamento de F\'{\i}sica, Instituto Tecnol\'{o}gico da Aeron\'{a}%
utica, CTA,

12228-900, S\~{a}o Jos\'{e} dos Campos, SP, Brazil\medskip

$^{3}$Instituto de F\'{\i}sica, Universidade Federal do Rio de Janeiro,

21941-972, Rio de Janeiro, RJ, Brazil\bigskip

The excitation of multiphonon giant resonances with heavy ions is discussed.
The conventional theory, based on the use of the virtual photon number
method in conjunction with the harmonic model is presented and its
shortcomings are discussed. The recently developed model that invoke the
Brink-Axel mechanism as an important contribution to the cross-section is
discussed and compared to the conventional, harmonic model. The decay
properties of these multiple giant resonances are also discussed within the
same coherent + fluctuation model in conjunction with the hybrid decay
model. It is demonstrated that the Brink-Axel mechanism enhances the direct
decay of the states, as data seem to require. Comparison of our model with
other recent theoretical works is presented.\bigskip

\noindent {\bf 1. INTRODUCTION}\smallskip

The study of the double giant dipole resonance in nuclei has received a
considerable amount of attention over the last 15 years [1]. Both the pion
double charge exchange and relativistic heavy ion Coulomb excitation
reactions have been used to probe this large amplitude collective motion in
many fermion systems. The quest for the similar double plasmon resonance in
metallic clusters is underway [2]. Plans are also in progress to search for
the triple giant dipole resonance (TGDR) in nuclei [3]. It is clearly of
importance to supply theoretical estimates of the cross-section as well as
the different decay branching ratios of these exotic collective modes. This
is the purpose of the present paper. We use the recently developed coherent
plus incoherent excitation theory of Ref. [4] in conjunction with the hybrid
decay model of Dias-Hussein-Adhikari (DHA) of Ref. [5].

The existing models for the calculation of the excitation cross-section of
DGDR can be grouped into four categories: a microscopic structure model in
conjunction with second order Coulomb excitation perturbation theory [6], a
macroscopic, oscillator model in the Weis\"{a}cker-Williams approximation[7]
coupled channels Coulomb excitation theory [8] and finally the recently
developed average plus fluctuation model [4,9]. In this latter model the
average cross-section is calculated according to the theory developed in
[11], where the simple, double, etc. giant resonances are considered as
doorway states belonging to the\linebreak

\newpage \topmargin 6pt \textheight 230mm \textwidth 160mm

\noindent spectrum of a damped harmonic oscillator. The recent work of Gu
and Weidenm\"{u}ller [12], based on random matrix theory, lends full support
to our model [4,9].

In this review we describe our work based on the Brink-Axel mechanism. In
section 2 we describe the harmonic oscillator model and the multi-step
reaction theory. In section 3 we consider the static and dynamic effects
associated with the damping of the multiphonon resonances. In section 4 we
present the ``Absorption-Fluctuation Theorem (AFT)''. In section 5 the
Brink-Axel mechanism is invoked to implement the AFT. In section 6 the
evolution equation of the density matrix is presented and discussed.
Application of the theory to the excitation of the DGDR and TGDR \ in
several systems is presented in section 7. In section 8 the decay of these
resonances is discussed within the hybrid model of DHA [5]. Finally, in
section 9 concluding remarks are given.\bigskip

\noindent {\bf 2. MULTISTEP REACTION THEORY\smallskip }

In this section we describre the semiclassical excitation of a harmonic
oscillator ,which represents the spectrum of the giant resonances. The
spectrum of an harmonic oscillator is given by, 
\begin{equation}
E_{n}=E_{0}+n\hbar \omega ,
\end{equation}%
\noindent where $E_{0}$ is the zero point energy, $n$ is the number of
phonons and $\omega $ is the frequency.

We now allow the harmonic oscillator to be excited by an external pulse $%
V(t) $ (a passing fast ion). The excitation, for zero width, is described by
the Poisson formula for the probabilities, vis.

\begin{equation}
P_{n}=\frac{\left[ P_{1}^{\left( 0\right) }\right] ^{n}}{n!}P_{0}.
\end{equation}%
Above, $n$ is the number of phonons in the excited state, $P_{1}^{\left(
0\right) }$ is the first order perturbation probability of exciting a one
phonon state, 
\begin{equation}
P_{1}^{\left( 0\right) }=\left\vert \frac{1}{i\hbar }\int\limits_{-\infty
}^{\infty }V(t)\,e^{i\omega t}dt\,\right\vert ^{2},
\end{equation}%
\noindent and $P_{0}$ is the probability for the oscillator to remain in the
ground state,

\begin{equation}
P_{0}=\exp \left[ -P_{1}^{\left( 0\right) }\right] .
\end{equation}

Thus for the double giant resonance (DGR) excitation one has,

\begin{equation}
P_{2}=\frac{1}{2}P_{1}^{\left( 0\right) }P_{1}^{\left( 0\right) }\exp \left[
-P_{1}^{\left( 0\right) }\right] \,.
\end{equation}

For the triple giant resonance (TGR), one has,

\begin{equation}
P_{3}=\frac{1}{6}P_{1}^{\left( 0\right) }P_{1}^{\left( 0\right)
}P_{1}^{\left( 0\right) }\exp \left[ -P_{1}^{\left( 0\right) }\right] \,,
\end{equation}

\noindent or

\begin{equation}
P_{3}=\frac{1}{3}P_{1}^{\left( 0\right) }P_{2}\,.
\end{equation}

In all of the above, the cross section $d\sigma _{n}\diagup d\Omega $ is
calculated using the semi-classical formula 
\begin{equation}
\frac{d\sigma _{n}}{d\Omega }=\frac{d\sigma _{Ruth}}{d\Omega }P_{n},
\end{equation}

\noindent and the integrated cross section is simply 
\begin{equation}
\sigma _{n}=\int \frac{d\sigma _{Ruth}}{d\Omega }P_{n}\left( b\right)
d\Omega =2\pi \int\limits_{0}^{\infty }bdbP_{n}\left( b\right)
\end{equation}%
\bigskip

{\bf \noindent 3. EFFECT OF DAMPING\smallskip }

Giant resonances are damped because of their coupling to the complicated
fine structure states and to the open decay channels. By far the former type
of damping is the dominant one. Thus the harmonic oscillator model must be
damped by including an appropriate damping width. The excitation energy of
the n-phonon state of the damped oscillator is given by $E_{n}-i\Gamma
_{n}/2,$ where $E_{n}=nE_{1}$ and $E_{1}$ is the excitation energy of the
single giant resonance. From the Boson nature of the phonons, one whould
expect, 
\begin{equation}
\Gamma _{n}=n\,\Gamma _{1}\,.
\end{equation}

Damping contains escape (to open channels) plus spreading (to complex
chaotic states). According to sum rule arguments\ [12] the widths of
multiple giant resonances should behave as $\Gamma _{n}=\sqrt{n}~\Gamma _{1}$%
. The experimental width is situated between the two limits. This fact
points to the conclusion that the experimentally determined width is an
effective one which may embody some reaction dynamic effects [9].

There is no closed form Poisson-type solution for the excitation
probabilities of the damped harmonic oscillator above. However, one can
solve the problem numerically, and as expected the DGR cross section is
reduced at high energies. At low energies there is a slight increase in the
cross section as a function of $\Gamma _{1}$, contrary to the claim of ref
[14\ ]. The reason is that in the absense of the width the number of virtual
photons for low bambarding energies at the position of the resonance is very
small, resulting in a small cross section.

\bigskip 

\begin{figure}[htb]
\begin{center}
\epsfig{file=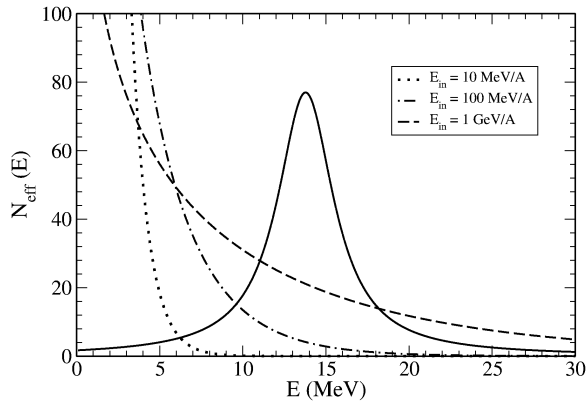, height=5.5 cm}
\caption{Effective virtual photon spectra at 
10, 100 and 1000 MeV/A together with the Breit-Wigner approximation to the
GDR spectral function of $^{208}$Pb.}
\end{center}
\end{figure}
\bigskip 

We should emphasize that the Poisson formula (or harmonic formula), Eq. (3),
can still be used in a microscopic calculation where all fine structure
states (that give rise to $\Gamma _{1}$) are included. This fact was nicely
demonstrated in Ref. [10] using a schematic microscopic picture containing a
simple harmonic oscillator, representing the giant resonance spectrum,
coupled to a ``bath'' of many simple harmonic oscillators. The excitation
probabilities were evaluated in closed form and were found to be 
\begin{equation}
P_{n}=\sum\limits_{\{\nu \}}P_{n}^{(\{\nu \})}=\exp \left( -\sum\limits_{\nu
\prime }\left| \alpha _{\nu \prime }\right| ^{2}\right) \frac{%
\sum\limits_{\nu }\left| \alpha _{\nu }\right| ^{2}}{n!}\,,
\end{equation}

\noindent where $\left| \alpha _{\nu }\right| ^{2}$ is the many-harmonic
oscillator generalization of $P_{1}^{\left( 0\right) }$ of Eq.(4), Ref. [10].

Short of the microscopic picture, one ``suggests'' the validity of the
harmonic, $P_{2}=\frac{1}{2}P_{1}^{\left( 0\right) }P_{1}^{\left( 0\right)
}\exp \left[ -P_{1}^{\left( 0\right) }\right] $, form for the DGR excitation
probability within the damped oscillator model. This is the widely used
folding model [15] whose results are employed in the analysis of the data.
It is referred to in the literature erroneously as the ``harmonic'' model.
It should be called the ``damped harmonic'' model.

The data analyses show [1] that the folding model does not fully account for
the data (to within $20{\%}$) and in the exceptional case of $Xe+Pb$ at the
GSI energies (640 MeV/A) the discrepancy is more than $80{\%}$. The reason
can be easily the contribution of dynamical effects that accompany the
damping, the Brink-Axel effect to be discussed below, as well as higher
order anharmonic effects of the type discussed in [16]. We should mention
that the dynamic effects become quite significant at lower energies, and it
would be important to fully test it with new data taken at, say, 100 MeV/A .

The data also seem to suggest that the direct decay of multiphonon states is
enhanced in comparison with the harmonic picture of the independent decay of
the multiphonons found in the state.

The above two features found in the data clearly call for the development of
a consistent reaction theory that accounts for both the excitation and the
decay of multi-phonon states in heavy-ion reactions. One would rely on the
idea of calculating an average amplitude (damped oscillator) that supplies
the coherent, harmonic (folding) contribution to the cross section. The
remaining piece of the amplitude, whose energy average is zero, supplies the
statistical, fluctuation contribution to the cross section ( the Brink-Axel
contribution), just like in low energy neutron scattering from nuclei where
one has the optical potential supplying the average amplitude and thus the
``direct'' cross section and the statistical amplitude that gives rise to
the fluctuation cross section calculated from the Hauser-Feshbach
theory.\bigskip

\noindent {\bf 4. ABSORPTION-FLUCTUATION THEOREM\smallskip }

In Nuclear Reaction Theory one uses average amplitudes to calculate optical
or coherent cross section. Unitarity dictates that one must take into
account a fluctuation or statistical cross section and add it incoherently
to the optical one to obtain the average cross section to be compared with
the data. There are three types of fluctuations.

- Initial state fluctuations: these occur in low energy nuclear reactions
that involve the formation of the compound nucleus, $N+A\rightarrow
CN\rightarrow N^{\prime }+A^{\prime }$, the S- matrix is given by: $S=\bar{S}%
+V\,G_{CN}$ $V$ : with $\bar{G}_{CN}=0$, where the upper bar implies energy
average. The resulting cross section contains an optical model piece plus a
fluctuation, Hauser-Feshbach one.

- Final state fluctuations: There are cases that involve fluctuations in the
final state of the reaction. This happens when the final channel contains a
particle in an unbound state (continuum). An example of this type of
fluctuations is the reaction $A\left( p,\gamma \right) $ at $E_{p}\sim 20MeV$%
. One gets a direct-semidirect cross-section and a direct-compound
(fluctuation) one. The latter was found to be dominant [17].

- Intermediate state fluctuations: These fluctuations occur in multistep
direct reactions. They involve fluctuation in the excited nucleus
(subsystem) in, e.g., a heavy-ion reaction, such as the excitation of the
DGDR: $GS\rightarrow GDR\rightarrow DGDR$. The S-matrix can be represented
by $S=\bar{S}+\Gamma ^{\downarrow }~G~V$, with $\bar{G}=0$. The
corresponding cross-section can be written as $\sigma =\sigma _{c}+\sigma
_{fl}$.\bigskip 

\noindent {\bf 5. THE BRINK-AXEL MECHANISM\smallskip }

Another important feature arising from the damping is the dynamical effect
owing to the existence of a mechanism that allows for the excitation of a
collective phonon on top of the background of chaotic compound states to
which the one phonon state decays in a multistep reaction proceeding
sequentially. The Brink--Axel phonons contribute incoherently to the cross
section since they correspond to a slower process compared to the sequential
excitation of the damped oscillator states. In fact one can calculate the
bombarding energies at which theses contribution become significant by
dividing the one-phonon decay time, $\tau _{d},$ by the collision time, $%
\tau _{c}$, vis. 
\begin{equation}
\frac{\tau _{d}}{\tau _{c}}=\frac{\hbar /\Gamma _{1}}{2b/\gamma v}.
\end{equation}

\noindent For $^{208}Pb$ with $\Gamma _{1}\approx 4MeV$, $b=15fm$ at $%
E=300MeV/A$, the above ratio is about $0.6$ and increasing as the energy is
lowered, indicating the increased competition between the one phonon decay
into the background and continuation of the sequential excitation. Therefore
one may envisage the following picture of the reaction dynamic.

\begin{figure}[tbh]
\begin{center}
\epsfig{file=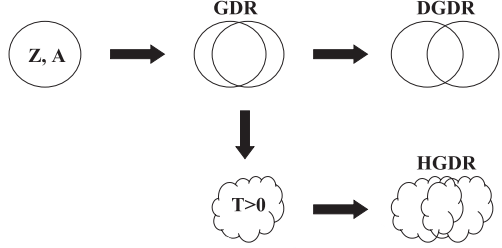, height=4.0 cm}
\end{center}
\caption{Cartoon depiction of the conventional double giant dipole resonance
excitation (DGDR) and the alternative `hot' giant dipole excitation (HGDR)
discussed here.}
\end{figure}
\bigskip

\noindent {\bf 6. EVOLUTION EQUATION\smallskip }

The equation that governs the evolution of the density matrix ,from which
the excitation probabilities are calculated is obtained from the damped
oscillater equation. The space used to derive the evolution equation is
depicted in Figure 3.

\begin{figure}[tbh]
\begin{center}
\epsfig{file=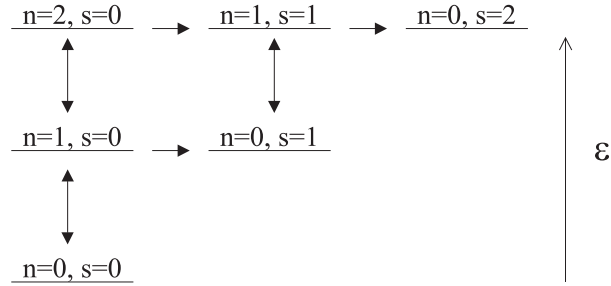, height=4.0 cm}
\end{center}
\caption{Schematic representation of the collective/statistical states and
their transitions. The vertical arrows represent the two-way coherent
excitation/de-excitation of collective phonons. The horizontal arrows
represent the one-way statistical decay of the collective phonons. $\protect%
\varepsilon $ denotes the excitation energy. }
\end{figure}

The time evolution equation for the density matrix was derived originally by
Ko [18] and in the present context by Carlson at al. [4] 
\begin{eqnarray}
\hbar \frac{\partial \rho _{nn^{\prime }}^{s}}{\partial t}
&=&-i\sum_{m}\left( \tilde{V}_{nm}(t)\rho _{mn^{\prime }}^{s}-\rho _{nm}^{s}%
\tilde{V}_{mn^{\prime }}(t)\right)  \label{evoleq} \\
&&-\frac{(\Gamma _{ns}+\Gamma _{n^{\prime }s})}{2}\rho _{nn^{\prime
}}^{s}+\delta _{nn^{\prime }}\sum_{r,m}\Gamma _{ns,mr}\rho _{mm}^{r}, 
\nonumber
\end{eqnarray}
The second line of Eq.(13) contains the incoherent contributions of the
statistical loss and gain terms, respectively.

Assuming that the collective excited states are harmonic $n$-phonon giant
dipole states, the interaction matrix elements can be written as 
\begin{equation}
\tilde{V}_{nn^{\prime }}(t)=\left( \exp \left[ i\varepsilon _{d}t/\hbar %
\right] \sqrt{n}\delta _{n^{\prime },n-1}+\exp \left[ -i\varepsilon
_{d}t/\hbar \right] \sqrt{n+1}\delta _{n^{\prime },n+1}\right) V_{01}(t)
\end{equation}
where $\varepsilon _{d}$ is the excitation energy of the giant dipole
resonance and $V_{01}(t)$ is the semiclassical matrix element coupling the
ground state to the giant resonance, which we take to have the simple form 
\begin{equation}
V_{01}(t)=V_{0}\frac{\left( b_{\min }/b\right) ^{2}}{1+(\gamma vt/b)^{2}},
\end{equation}
as given in Ref.~[11]. As is done there, we neglect the spin degeneracies
and magnetic multiplicities of the giant resonance states and approximate
the projectile-target relative motion as a straight line.

In the case of harmonic phonons, the decay widths can be approximated as 
\begin{equation}
\Gamma _{ns}=n\Gamma _{1},
\end{equation}%
where $\Gamma _{1}$ is the spreading width of the giant dipole resonance. We
have neglected the contribution to the width of the hot statistical
background of states since, at the low temperatures involved here, the decay
widths of the hot Brink-Axel resonances are very similar to those of the
cold ones.

We have performed calculations of multiple giant dipole resonance excitation
within the model for the system $^{208}$Pb + $^{208}$Pb in the projectile
energy range from 100 to 1000 Mev/nucleon. In Fig.~4, we show the
differential excitation cross section that we obtain for the system $^{208}$%
Pb + $^{208}$Pb at 640 MeV/nucleon as a function of the excitation energy.
This was obtained by summing Breit-Wigner expressions with the appropriate
excitation energy and width for each of the total $n$-phonon cross sections.
We show only the contributions of the first three giant dipole resonances,
as the higher order ones are almost invisible even on our theoretical curve.
Only the first and second giant dipole resonances have been observed
experimentally.

\begin{figure}[htb]
\begin{center}
\epsfig{file=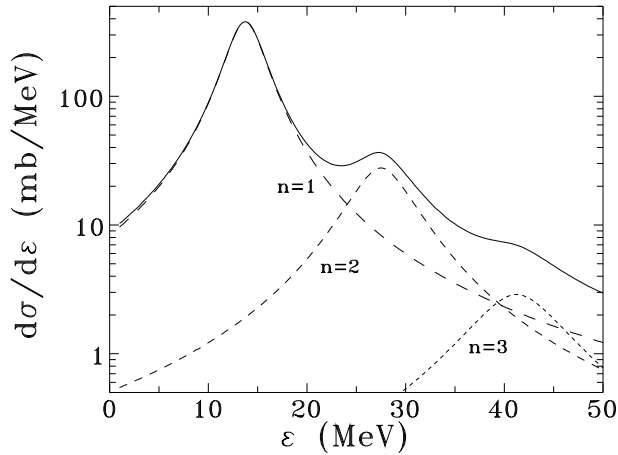, height=6.0 cm}
\caption{Theoretical multiple giant resonance differential excitation cross
section of $^{208}$Pb at a projectile energy of 640 MeV/nucleon.}
\end{center}
\end{figure}

\bigskip

{\bf \noindent 7. DGDR AND TGDR EXCITATION\smallskip }

To be consistent with the discussion of the AFT of section 4, it seems
appropriate to write the cross-section for the excitation of the DGDR and
the TGDR as, respectively,

\begin{equation}
\sigma ^{\left( 2\right) }=\sigma _{c}^{\left( 2\right) }+\sigma
_{fl}^{\left( 2\right) }\left( 1\right)
\end{equation}

\begin{equation}
\sigma ^{\left( 3\right) }=\sigma _{c}^{\left( 3\right) }+\sigma
_{fl}^{\left( 3\right) }\left( 2\right) +\sigma _{fl}^{\left( 3\right)
}\left( 1\right) \,,
\end{equation}

\noindent where $\sigma _{fl}^{\left( n\right) }\left( i\right) $ denotes
the contribution of $i$ Brink-Axel phonons to the excitation cross-section
of $n$ giant resonances phonons [19].

We have calculated the excitation cross-sections, $\sigma ^{(1)}$, $\sigma
^{(2)}$ and $\sigma ^{(3)}$, for various nuclei incident on $^{208}$Pb at
several bombarding energies, using a three-dimensional (3D) generalization
of the model of Ref. [4]. The 3D time evolution equation used in Ref. [19]
to describe the excitation and decay of the GDR phonons possesses the same
form as the one-dimensional equation of Ref. [4]. However, the collective
and statistical excited states of the 3D model take into account all
possible combinations of the (two) transverse and (one) longitudinal degrees
of freedom, which yield 3 coherent one-phonon states, 6 coherent two-phonon
states and 10 coherent three-phonon states as well as a multitude of states
containing a mixture of coherent and statistical excitations. Decays of the
three types of phonons to the statistical background are assumed to occur
independently but to each obey Bose-Einstein statistics.

The Coulomb interaction matrix elements used to describe the transverse
modes of the GDR excitation in the 3D model are the physically appropriate
ones, as given in Ref. [11]. The longitudinal Coulomb interaction matrix
element, however, is modified from the form given there. Through a guage
transformation, it is reduced to a term proportional to the longitudinal
component of the eletric field, in analogy to the transerse terms, but which
differs from the expression given in Ref. [11] by a total time derivative.
We emphasize that our theory contains the effect of the adiabaticity to all
orders. If we neglect the width of the GDR and use perturabtion theory, we
fully recover the model of Baur and Bertulani[7].

As in Ref. [4], the coupled equations of motion are solved as a function of
impact parameter to yield asymptotic occupation probabilities. Effective
asymptotic occupation probabilities are defined, for states that decay, as
the sum over the probability that decays out of each state during the time
evolution. Cross sections are obtained by integrating each probability x
differential area over impact parameter and summing over polarizations.

The various contributions to the cross sections are easily extracted from
the theoretical calculations. In Table I, we present the coherent and
fluctuation contributions to the DGDR cross section, $\sigma _{c}^{(2)}$ and 
$\sigma _{fl}^{(2)}(1)$ for various nuclei incident on $^{208}$Pb at several
energies. We use a global systematic for the GDR energies and widths: $%
E_{GDR}=43.4A^{-.215}$ MeV and $\Gamma _{GDR}=0.3E_{GDR}$[4]. The energies
of the DGDR and TGDR resonances were taken to be two and three times those
of the GDR, respectively, since simple anharmonic effects are small [20].
The widths of the DGDR and TDGDR were taken in Ref. [19] to be $\sqrt{2}$
and $\sqrt{3}$ times those of the GDR widths, respectively, as rigorous sum
rule arguments indicate [13].

\begin{table}[tbp]
\caption{ Contributions of the coherent and fluctuation components to the
DGDR excitation cross section (in mb) of various projectiles incident on a
lead target at two values of the incident energy.}
\par
\begin{center}
\begin{tabular}{|c|rrr|rrr|}
\hline
& \multicolumn{3}{c|}{100 MeV} & \multicolumn{3}{c|}{1 GeV} \\ 
Projectile & $\sigma_{c}^{(2)}$ & $\sigma_{fl}^{(2)}(1)$ & $%
\sigma_{tot}^{(2)}$ & $\sigma_{c}^{(2)}$ & $\sigma_{fl}^{(2)}(1)$ & $%
\sigma_{tot}^{(2)}$ \\ \hline
$^{40}$Ca & 2.17 & 2.19 & 4.36 & 7.20 & 0.72 & 7.92 \\ 
$^{120}$Sn & 26.48 & 22.94 & 49.42 & 72.61 & 6.65 & 79.26 \\ 
$^{132}$Xe & 32.19 & 27.57 & 59.76 & 88.50 & 8.00 & 96.50 \\ 
$^{165}$Ho & 51.13 & 42.60 & 93.73 & 138.59 & 12.34 & 150.93 \\ 
$^{208}$Pb & 96.95 & 72.87 & 169.82 & 234.84 & 19.83 & 254.67 \\ 
$^{238}$U & 109.15 & 84.86 & 194.01 & 276.53 & 24.04 & 300.59 \\ \hline
\end{tabular}%
\end{center}
\end{table}

In Table II, we present the contributions to the TGDR cross section, $\sigma
_{c}^{(3)}$, $\sigma _{fl}^{(3)}(2)$, and $\sigma _{fl}^{(3)}(1)$. We
observe that the cross sections increase dramatically with the charge of the
projectile. As is well known, the coherent two-phonon cross sections scales
approximately as the projectile charge Z squared, while the three-phonon one
scales as $Z^{3}$. We also observe that the coherent contribution to the
cross sections only dominates at relatively high incident energies. At $%
E/A=100$ MeV, it is clear from the tables that the fluctuation contribution
to the DGDR cross section is about as large as the coherent one, while the
fluctuation contribution to the TGDR is about three times larger than the
coherent contribution.

\begin{table}[tbp]
\caption{ Contributions of the coherent and fluctuation components to the
TGDR excitation cross section (in mb) of various projectiles incident on a
lead target at two values of the incident energy.}
\begin{tabular}{|c|rrrr|rrrr|}
\hline
& \multicolumn{4}{c|}{100 MeV} & \multicolumn{4}{c|}{1 GeV} \\ 
Projectile & $\sigma_{c}^{(3)}$ & $\sigma_{fl}^{(3)}(2)$ & $%
\sigma_{fl}^{(3)}(1)$ & $\sigma_{tot}^{(3)}$ & $\sigma_{c}^{(3)}$ & $%
\sigma_{fl}^{(3)}(2)$ & $\sigma_{fl}^{(3)}(1)$ & $\sigma_{tot}^{(3)}$ \\ 
\hline
$^{40}$Ca & 0.02 & 0.06 & 0.02 & 0.10 & 0.11 & 0.02 & 0.00 & 0.13 \\ 
$^{120}$Sn & 0.84 & 1.92 & 0.64 & 3.40 & 3.03 & 0.47 & 0.04 & 3.54 \\ 
$^{132}$Xe & 1.10 & 2.50 & 0.83 & 4.43 & 4.07 & 0.62 & 0.05 & 4.74 \\ 
$^{165}$Ho & 2.08 & 4.70 & 1.54 & 8.32 & 7.76 & 1.17 & 0.09 & 9.02 \\ 
$^{208}$Pb & 5.28 & 10.78 & 3.36 & 19.42 & 16.68 & 2.40 & 0.18 & 19.26 \\ 
$^{238}$U & 6.13 & 13.14 & 4.22 & 23.49 & 21.01 & 3.14 & 0.24 & 24.39 \\ 
\hline
\end{tabular}%
\end{table}

The reason for the unexpected larger cross sections at lower energies can be
traced to two factors. The average cross section is larger than the that of
Baur and Bertulani owing to the fact that the inclusion of the width of the
one-phonon resonance (GDR) allows the excitation of that resonance, as well
as the DGDR and TGDR, even at very low excitation energies, where the
virtual photon spectrum is concentrated at low bombarding energy. This
enhancement can easily be missed, if the width is not taken into account, as
in the original Weis\"{a}cker-Williams approximation[7]. The second reason
for the increase in the DGDR and the TGDR cross sections, which is also
related to the inclusion of the width, is the Brink-Axel fluctuation
contribution, which tends to increase as the bombarding energy is
lowered.\bigskip

\noindent {\bf 8. DECAY OF MULTIPLE GIANT RESONANCES\smallskip }

We turn now to the decay of the DGR and TGR. We first remind the reader of
the hybrid direct+fluctuation decay model of DHA [5]. According to this
model, which has been extensively used in the analysis of decay data
[21,22], the GR decays to a find channel $f$ in the following manner:

\begin{eqnarray}
\sigma _{f}^{(1)} &=&\sigma ^{(1)}\left[ \left( 1-\mu _{1}\right) \frac{\tau
_{f}^{(GR)}}{\sum_{j}\,\tau _{j}^{(GR)}}+\mu _{1}\frac{\tau _{f}^{(CN)}+\mu
_{1}\tau _{f}^{(GR)}}{\sum_{j}\,\left( \tau _{j}^{(CN)}+\mu _{1}\tau
_{j}^{(GR)}\right) }\right]  \nonumber \\
&\equiv &\sigma ^{(1)}\,(P_{f}^{\uparrow }+P_{f}^{\downarrow })\,,
\label{grdec}
\end{eqnarray}
\noindent where $\sigma ^{(1)}$ is the one phonon excitation cross section
discussed before, while $\mu _{1}=\frac{\Gamma _{1}^{\downarrow }}{\Gamma
_{1}}$ and the $\tau $'$s$ are the appropriate transmission coefficients. We
have written the probability of populating the final channel $f$ through
direct decay of the GR as 
\begin{equation}
P_{f}^{\uparrow }=\left( 1-\mu _{1}\right) \frac{\tau _{f}^{(GR)}}{%
\sum_{j}\,\tau _{j}^{(GR)}}\,,
\end{equation}
\noindent and the probability of of populating the channel $f$ through the
statistical states as 
\begin{equation}
P_{f}^{\downarrow }=\mu _{1}\frac{\tau _{f}^{(CN)}+\mu _{1}\tau _{f}^{(GR)}}{%
\sum_{j}\,\left( \tau _{j}^{(CN)}+\mu _{1}\tau _{j}^{(GR)}\right) }
\end{equation}
\noindent Note that the statistical decay component contains explicit
reference to the GR direct transmission, $\left( \mu _{1}\tau
_{f}^{(GR)}\right) $.

A detailed derivation of the direct decay of the DGDR and TGDR was presented
in Ref. [19]. Here we only present the final result, 
\begin{equation}
\frac{\sigma _{f}^{(2)}}{\sigma _{c,f}^{(2)}}=1+\frac{1}{2}\frac{\sigma
_{fl}^{(2)}(1)}{\sigma _{c}^{(2)}}\,,
\end{equation}%
and 
\begin{equation}
\frac{\sigma _{f}^{(3)}}{\sigma _{c,f}^{(3)}}=1+\frac{2}{3}\frac{\sigma
_{fl}^{(3)}(2)}{\sigma _{c}^{(3)}}+\frac{1}{3}\frac{\sigma _{fl}^{(3)}(1)}{%
\sigma _{c}^{(3)}}\,.
\end{equation}%
\noindent Thus, a considerably larger direct decay may occur if the
fluctuation contributions are important, which may occur at lower bombarding
energies. Of course, one could obtain deviation of the direct decay from the
harmonic limit (two or three independently decaying phonons), if anharmonic
effects were allowed, This, however, will imply deviation of the spectrum of
the oscillator from the harmonic sequence, which seems to be borne out
neither by experiment [1] nor by calculation [20], notwithstanding the
higher order anharmonic effects considered in Ref. [16].\bigskip

{\bf \noindent 9. CONCLUSIONS\smallskip }

We have developed a description of multiple giant resonance excitation and
decay that incorporates incoherent (statistical) contributions to the cross
section. The incoherent contributions arise from the excitation of
collective phonons on top of the statistical background that results from
the decay of previously excited collective phonons, namely, the Brink-Axel
mechanism. Semi-classical calculations show that incoherent, B-A,
contributions are an important part of the cross section at low to
intermediate energies. These contributions are expected to be important in
all regions of the mass table. This phenomenon is not limited to the
excitation of collective modes in nuclei. It is of potential importance in
any system in which multiply excited collective modes couple and decay to a
statistical background of states. The decay of multi-phonon giant resonances
is discussed within the hybrid, direct + statistical, model of Dias-
Hussein- Adhikari (DHA). It is found that the B-A mechanism enhances the
direct decay component of multi-phonon giant resonances.\bigskip

\noindent {\bf ACKNOWLEDGMENT\medskip }

This work was supported in part by FAPESP and the CNPq.\bigskip

\noindent {\bf REFERENCES\medskip }

\begin{enumerate}
\item See, e.g., T. Aumann, P. F. Bortignon and H. Emling, Ann. Rev. Nucl.
Part. Sci. 48 (1998) 351.

\item B.V. Issendorff, Talk given at DGDR99 ECT*- Trento, Italy, May 1999.

\item N. Frascaria, Talk given at DGDR99 ECT* - Trento, Italy, May 1999;
Nucl. Phys. A687, 154c (2001), (private communication).

\item B.V. Carlson, M.S. Hussein, A.F.R. de Toledo Piza and L.F. Canto,
Phys. Rev. C60, 014604 (1999); B.V. Carlson and M.S. Hussein, Phys Rev. C59,
R2343 (1999).

\item H. Dias, M.S. Hussein and S.K. Adhikari, Phys. Rev. Lett 57, 1988
(1986); R. Bonetti, M. Chadwick, P. Hodgson, B.V. Carlson and M.S. Hussein,
Phys. Rep. 202, 171 (1991).

\item V. Yu Ponomarev, E. Vigezzi, P.F. Bortignon, R.A. Broglia, G. Colo, G.
Lazzari, V.V. Voronov and G. Baur, Phys. Rev. Lett. 72, 1168 (1994).

\item G. Baur and C.A. Bertulani, Phys. Lett. B174, 28 (1986).

\item See, e.g., C.A. Bertulani and V. Yu Ponomaev, Phys. Rep. 321 139
(1999).

\item B.V Carlson, L.F. Canto, S. Cruz-Barrios, M.S. Hussein and A.F.R. de
Toledo Piza, Ann. Phys. (New York) 276 111 (1999); L.F. Canto, B.V. Carlson,
M.S. Hussein and A.F.R. de Toledo Piza, Phys. Rev. C60, 064624 (1999).

\item B.V Carlson, L.F. Canto, S. Cruz-Barrios, M.S. Hussein and A.F.R. de
Toledo Piza, Phys. Rev. C59 (1999) 2689.

\item L.F. Canto, A. Romanelli, M.S. Hussein and A.F.R. de Toledo Piza,
Phys. Rev. Lett. 72, 2147 (1994); C.A. Bertulani, L.F. Canto, M.S. Hussein
and A.F.R. de Toledo Piza, Phys. Rev. C53, 334 (1996).

\item J.Z. Gu and H.A. Weidenm\"{u}ller, Nucl. Phys. A690, 382 (2001).

\item E.J.V. Passos, M.S. Hussein, L.F. Canto, and B.V. Carlson, Phys. Rev.
C65, 034326 (2002); H. Kurasawa and T. Suzuki, Nucl. Phys. A597, 374 (1996).

\item G. Baur, C.A. Bertulani and D. Dolci, Eur. Phys. A7 (2000) 55.

\item W.J. Llope and P. Braun-Munzinger, Phys. Rev. C41 (1990) 2644

\item M. Fallot, Ph. Chomaz, M.V. Andres, F. Catara, E.G. Lanza and J.A.
Scarpaci, submitted to Nucl. Phys. A.

\item W.E. Parker et al. , Phys. Rev.C 52 (1995{\it ) 252}

\item C.M. Ko, Z. Phys. A286 (1978) 405.

\item M.S. Hussein, B.V. Carlson, L.F. Canto and A.F.R. de Toledo Piza,
Phys. Rev. C66 (2002) 034615.

\item V. Yu. Ponomarev, P.F. Bortignon, R.A. Broglia, and V.V. Voronev,
Phys. Rev. Lett. 85, 1400 (2000); D.T. de Paula, T. Aumann, L.F. Canto, B.V.
Carlson, H. Emling, M.S. Hussein, Phys. Rev. C64, 064605 (2001).

\item A. Bracco, J.R. Beene, N. Van Giai, P.F. Bortignon, F. Zardi and R.A.
Broglia, Phys. Rev. Lett. 60, 2603 (1988); A. Bracco, P.F. Bortignon and
R.A. Broglia, ``Giant Resonances'', 1996.

\item B.V. Carlson, M.S. Hussein, and A.F.R. de Toledo Piza, Phys. Lett.
431B, 249 (1998).
\end{enumerate}

\end{document}